\documentclass{JHEP3}
\usepackage{epsfig}
\newcommand{\be}{\begin{equation}}
\newcommand{\ee}{\end{equation}}
\newcommand{\bea}{\begin{eqnarray}}
\newcommand{\eea}{\end{eqnarray}}
\newcommand{\IR}{\mathbb{R}} 

\def\IZ{\relax\ifmmode\hbox{Z\kern-.4em Z}\else{Z\kern-.4em Z}\fi}
\newcommand{\IS}{{\bf S}}

\newcommand{\non}{\nonumber \\}

\def\half{{1 \over 2}} 

\def\del{{\partial}}
\def\room{~\rule[-2mm]{0mm}{8mm}}

\def\hG{{\hat G}} \def\hA{{\hat A}}
\def\hA{\hat{A}}  

\def\cL{{\cal L}}

\def\wttriangle{\widetilde{\triangle}}
\def\tG{\tilde{G}} \def\tx{\tilde{x}}
\def\tQ{\widetilde{Q}}

\def\room{~\rule[-2mm]{0mm}{8mm}}
\def\presub{\vspace{.5cm} \noindent}

\def\bi{\begin{itemize}} \def\ei{\end{itemize}}

\def\Schw{Schwarzschild }
\def\({\left(} \def\){\right)}
\def\[{\left[} \def\]{\right]}
\preprint{{\tt hep-th/0608001}}

\title{ \center{A New Action-Derived Form of The Black Hole Metric}}

\author{
Barak Kol \\
 Racah Institute of Physics\\
 Hebrew University \\
 Jerusalem 91904,
 Israel\\
{\tt barak\_kol@phys.huji.ac.il}}

\abstract{Several different but equivalent forms of the spherical
black hole (Schwarzschild) are known, as well as several
derivations. Here a novel form is derived through the power of the
action formalism. The method generalizes to any spherical black hole
or brane, and the charged hole case (Reissner-Nordstr\"{o}m) is
worked out in detail.}

\begin{document}

\begin{flushright}
\emph{To Eitan Aharoni,  \hspace*{2.7cm} \\
 my friend ever since the 7th grade, \\
 who always encourages me \hspace*{1.3cm} \\
 to examine fundamental issues}. \hspace*{0.5cm}
\end{flushright}


\noindent {\bf Introduction}.
 In 1916 Schwarzschild \cite{Schw}
discovered the metric which bears his name, and in perspective it
is recognized as the first \emph{black hole}. It is static,
spherically symmetric and uncharged.

By now several equivalent analytical metrics are known, all
describing the same black hole, while each one highlights different
properties. Each such metric is usually derived by choosing a gauge
for the metric, writing the Einstein equations and then solving
them. In general, while a given result may be proven in many ways,
some scientists believe that ``somewhere'' there exists a ``book''
with ``{\it the} proof''.  In this paper we shall describe such a
unique derivation for the metric of spherical black holes.

Very generally in physics it is known that an action, when
available, is the most concise packaging of the equations of motion,
and moreover it enables the widest class of field transformations.
Still in General Relativity (GR) traditionally one writes down the
equations of motion, even though an action principle is
known\footnote{Namely, the Einstein-Hilbert action supplemented in
the case of a fixed boundary by the York-Gibbons-Hawking term
\cite{York,GibbonsHawking}.}. The reason for that is the large gauge
symmetry: in order for the action to encode all the equations, it
must be written as a function of as many fields as there are
equations, while in GR, and in particular in deriving black hole
metrics, it is common practice to start by fixing the gauge (an
ansatz for the metric) in order to minimize the number of fields
involved and to simplify the equations.

Owing to its advantages, this paper takes the action approach to
derive the black hole metric. In order not to lose any of the
Einstein equations it is essential to keep a larger set of metric
components than in a ``most general'' ansatz, but not all of them
(in the presence of isometries), as discussed below in detail. This
has the benefit that the gauge can be chosen rationally rather than
arbitrarily, according to the form of the action. As we shall see
below, the gauge can be chosen according to the standard requirement
that the kinetic term be canonical. This in turn will simplify the
equations of motion and will lead to a new form of the black hole
metric.

\presub {\bf Ansatz}. Consider a spherical black hole in General
Relativity (so that the only field is the space-time metric) in
asymptotically flat $d$-dimensional space-time, namely the \Schw
black hole \cite{Schw} (see \cite{Tangherlini} for a generalization
to $d \neq 4$).

Which fields must be kept in an action approach? For a generic
metric we need to keep all the metric components $g_{\mu\nu}$ in
order to encode all the equations, namely all the components of
the Einstein tensor $G_{\mu\nu}$. However, in the common case of
\emph{a metric with isometries} a reduction in the number of
fields is possible. Given a metric which preserves the isometries,
its Einstein tensor is invariant as well, and in particular some
of its components necessarily vanish. Correspondingly fewer fields
are required.

For the case at hand, the spherical black hole, with its
$SO(d-1)_\Omega \times U(1)_t \times \IZ_{2,t}$ isometries, namely
spherical, stationary and time reflection, respectively, this
\emph{``maximally general ansatz''} is given by \footnote{Due to the
continuous isometries the fields cannot depend on $t$ nor on the
angular coordinates, $\Omega$. The $g_{\Omega i}$ components, where
$i \neq \Omega$, must vanish since there is no spherically symmetric
vector field on the sphere, while the $g_{tj}$ must vanish for all
$j \neq t$ due to time reflection combined with $t$-independence.}
\be
 ds^2=-e^{2\, A}\, dt^2 + e^{2\, B}\, dx^2 +e^{2\, C}\,
 d\Omega^2_{d-2} ~, \label{general-ansatz} \ee
 where $x$ denotes the radial coordinate (to distinguish it from
Schwarzschild's $r$),  and $d\Omega^2_{d-2}$ is the line element of
the $d-2$ sphere $\IS^{d-2}$, for example in 4d
$d\Omega^2=d\theta^2+\cos^2(\theta)\, d\phi^2$. At each point in
space-time the orbit of the symmetries generate a copy of the $d-1$
dimensional space $\IR_t \times \IS^{d-2}_\Omega$ and so the problem
is essentially 1d (the variable $x$). Accordingly, $A,B,C$ are
functions of $x$ alone.

We note that our ``maximally general ansatz'' is not the same as the
commonly used term ``the most general ansatz''. The latter usually
means that any metric can be put in that form, while our ansatz is
more general than that and is such that all of the Einstein
equations can be obtained by varying the gravitational action with
respect to its fields. Actually, if we fix the
$r$-reparameterization gauge in our ansatz and reduce the number of
fields to 2, it will still be ``most general''.

\presub {\bf Action.} Take the action to be minus the
Einstein-Hilbert action \be
 S=-S_{EH}=-{1 \over 16 \pi G}\, \int R\, dV =
  -{1 \over 16 \pi G}\, \int R\, \sqrt{-g}\, d^dx \ee
so that the kinetic terms will have the canonical sign even though
the variable $x$ is Euclidean. In order to express it in terms of
the fields in (\ref{general-ansatz}) we use \bea
 \sqrt{-g} &=& e^{A+B+(d-2)C} \non
 ds^2 &=& ds^2_X + \sum_i\, e^{2\, F_i(x)}\, ds^2_{Yi} \Rightarrow \non
 R &=& R_X + \sum_i\, [ e^{-2\, F_i}\, R_{Yi} - 2\, d_i\,
 \wttriangle (F_i) - d_i (\del F_i)^2] -\sum_{i,j}\, d_i\, d_j\,
 (\del F_i \cdot \del F_j) \label{Rfibration} \eea
 and after integration by parts we find \bea
 S_p &=& {\Omega_{d-2} \over 16 \pi G}\, \int dt\,  \int dx\,
 e^{A+B+(d-2)C} \cdot \non
 &\cdot& \[-e^{-2\,B} [2(d-2)\,A'\, C'+(d-2)(d-3)\,C'^2]
  -(d-2)(d-3)\, e^{-2\, C}\] \non
 \label{gauged-unfixed-action} \eea
 where $\Omega_{d-2}$ is the volume of the sphere $\IS^d$, the
factor $(d-2)(d-3)$ is its Ricci scalar, the subscript $p$ stands
for ``pre-gauge-fixing'' and from now on we shall omit the $x$
independent factor $\Omega_{d-2}/( 16 \pi G)\, \int
 dt$.

\noindent {\bf Fixing the gauge}. Note that as expected $B'$ does
not appear in the kinetic term, since $B$ represents the gauge
freedom of $x$ re-parametrization and can be set arbitrarily. At
this point we fix the gauge using the standard criterion of a
simplified kinetic term. By taking \be
 B=A+(d-2)C \label{gauge-choice} ~,\ee
 the kinetic coefficients (``the kinetic metric'') become $x$ independent.
The gauge-fixed action becomes \bea
 S &=& \int dx \( K_{ij}\, \del_x \phi^i \, \del_x \phi^j  - V \)\non
 \phi^i &:=& \[ \begin{array}{c}
  A \\ C  \end{array} \] \non
 K_{ij} &:=& -(d-2) \[ \begin{array}{cc}
  0 & 1 \\
  1 & (d-3) \end{array} \] \non
 V &:=& (d-2)(d-3)\, e^{2(A+(d-3)C)}  \label{action}
 ~,\eea
 and it should be supplemented by the constraint \be
 0=-\left. {\del S_p \over \del B} \right|_{B=A+(d-2)C} =
 K_{ij}\, \del_x \phi^i \, \del_x \phi^j  + V ~.\ee

\presub {\bf Solving the equations of motion}. Thus we have a
mechanical problem with two degrees of freedom and a kinetic term
which is non-positive-definite (the signature is actually $(1,1)$
since $\det (K)<0$). The potential depends only on a single linear
combination of the fields \be
 \hG:=A+(d-3)C ~.  \label{hG-def}\ee
Since $V=V(\hG)$ the two degrees of freedom decouple and the
direction orthogonal to $\hG$ (in the $K$-metric) will be free.
This direction turns out to be $A$ as can be seen from the
following diagonal form of $K$ \be
 K_{ij}\, \del_x \phi^i \, \del_x \phi^j
 = {d-2 \over d-3} \( A'^2-\hG'^2 \) ~.\ee
 Note that $\hG$ has a negative kinetic term.

Finally we can clean up some constants by pulling out of the
action the factor $(d-2)/(d-3)$ and defining \be
 G=\hG+\log(d-3) ~. \label{G-def} \ee
The final form for the action (and constraint) is \bea
 S &=& \int dx\, \[ A'^2-G'^2 - e^{2G}\] \label{final-action} \\
 0 &=& A'^2-G'^2 + e^{2G}
 \label{final-constr}
 ~. \eea
The resulting equations of motion are \bea
  A'' &=& 0 \non
  G'' &=& e^{2G} \label{eom} .\eea

Let us count the number of parameters for the solutions. Two second
order equations require 4 initial conditions, minus one (first
order) constraint leaves us with 3 initial conditions. Accounting
for the $x$ translation invariance of the system we are left with 2
parameters. This exactly matches our expectation for two black hole
parameters: $r_0$ and $V_\infty$ where $r_0$ is the \Schw radius and
$V_\infty$ is the potential at infinity (or $g_{tt}$ at infinity).
The latter is often not counted as a parameter since it can be
absorbed by a rescaling of $t$. On the other hand there are cases
where we do not allow for $t$ rescaling: it makes physical sense to
consider a black hole in a background where the asymptotic Newtonian
potential is non-zero. Thus we shall keep this parameter.

\begin{figure}[t!]
\centering \noindent
\includegraphics[width=10cm]{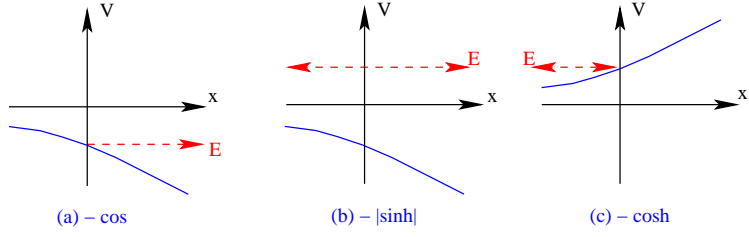}
\caption[]{The various cases for motion in one dimension with an
exponential potential, and the function $F$ appearing in their
solution (\ref{e-to-G}): (a) negative potential $V<0$ and negative
energy $E<0$ yield $F=\cos$, (b) $V<0,\, E>0$ yield $F=|\sinh|$, (c)
$V>0,\, E>0$ yield $F=\cosh$.} \label{exp-pot-figure}
\end{figure}

Since exponential potentials are common in gravity, it is
worthwhile to pause and determine the solution for a slightly more
general Lagrangian (with a single degree of freedom) \bea
 \cL &=& G'^2-V \non
 V &=& \pm e^{2G} ~,\eea
 allowing for either sign of the potential. As usual the energy is a
 constant of motion \be
 G'^2+ V =E :=\pm k^2 ~, \label{def-k} \ee
where the sign of the energy can be chosen independently (actually
there are only 3 possibilities since if $V>0$ then also $E>0$). $k$,
which we take to be positive, may be scaled out by \bea
 x &=& \tx/k \non
 G &=& \tG + \log(k) ~.\eea
The solutions are \be
 e^{-G}={1 \over k}\, F[k(x-x_0)] \label{e-to-G} ~,\ee
where the function $F$ is either $\cos,|\sinh|$ or $\cosh$ depending
on the three possible sign cases as shown in figure
\ref{exp-pot-figure}. Note that we have the most general solution
since it contains two constants $E$ and $x_0$ where $x_0$ simply
translates $x$. Note also that even functions correspond to motions
with a turning point, and that hyperbolic functions correspond to
motions which are asymptotically potential free.

In our case (\ref{eom}) the potential is negative\footnote{More
precisely, it has the sign opposite to $G$'s kinetic term.}, and
the energy is positive due to the constraint (\ref{final-constr}),
so we are in case (b) of the figure.
 The most general solution is \bea
 A &=& A_1\, x + A_0 \non
 e^{-G} &=&  \left|{1 \over A_1}\,  \sinh (A_1\, x) \right| ~,\eea
 where $x_0$ was set such that $e^{-G}$ vanishes at $x=0$, two other parameters
 $A_0,\, A_1$ are displayed, and without loss of generality $A_1$ will be
chosen to be positive from now on, just like $k$ was in
(\ref{def-k}).

Substituting this solution in the ansatz (\ref{general-ansatz})
using (\ref{gauge-choice},\ref{hG-def},\ref{G-def}) we arrive at
\emph{our final expression for the metric} \bea
 ds^2 &=& -e^{2A}\,dt^2+ e^{-2A/(d-3)}\, \[\exp\({2(d-2) \over d-3}\,\hG\)\, dx^2 +
   \exp\({2 \over d-3}\,\hG\)\, d\Omega^2_{d-2} \] =   \non
 &=& \room
   -e^{2 (A_1 x+A_0)}\, dt^2+ e^{-2 (A_1 x+A_0)/(d-3)} \cdot  \non
  &\cdot& \[ \({A_1 \over (d-3)\, \sinh(A_1 |x|)}\)^{2(d-2)
\over d-3} dx^2 +
 \({A_1 \over (d-3)\, \sinh(A_1 |x|)}\)^{2 \over  d-3} d\Omega^2_{d-2} \]
 \label{x-metric} ~.\non
  \eea
 The first line stresses that if we consider  a Kaluza-Klein
reduction over $t$, then the $d-1$ Weyl rescaled metric (the
Einstein frame) depends only on $\hG$ and not on $A$. The 4d case
is of special interest and the formulae simplify to \be
 ds^2 = -e^{2\, (A_1\, x+A_0)}\, dt^2+ e^{-2\, (A_1\, x+A_0)}\,
 \[ \({A_1 \over \sinh(A_1 |x|)}\)^4\,
 dx^2 + \({A_1 \over \sinh(A_1 |x|)}\)^2\,d\Omega^2_{d-2}\]
 ~. \label{4d} \ee

\presub {\bf Investigation of the metric}. Let us compare this
solution with the standard \Schw solution \bea
 ds^2= -f\, dt^2 + f^{-1}\, dr^2 + r^2\, d\Omega^2_{d-2} \non
 f=1-\({r_0 \over r}\)^{d-3} ~. \label{Schw}\eea
Since in this form $-g_{tt}|_{r \to \infty}=1$, and since $r \to
\infty$ will be seen shortly to correspond to $x \to 0$, we must set
$A_0=0$. Comparing $g_{\Omega \Omega}$ we find the change of
variable $r=r(x)$ and transformation of the parameter $r_0=r_0(A_1)$
to be \bea
 r^{d-3}&=& {2\, A_1 \over (d-3)(1-\exp(-2\, A_1\, |x|))} \label{x-to-r} \\
 r_0^{~d-3}&=& {2\, A_1 \over d-3} ~.  \label{A1-to-r0}\eea
Actually, one notices that the solution (\ref{x-metric}) depends on
$x$ only through the combination $y:=\exp(2\, A_1\, x)$ in terms of
which it becomes \bea
 ds^2 &=& -e^{2 A_0}\,y\, dt^2+ e^{-2 A_0/(d-3)}\, \cdot \label{y-metric} \\
 & \cdot& \[ \left({2\, A_1 \over (d-3)\, (1-y)} \right)^{2 \over
d-3}
  {dy^2 \over y\, (1-y)^2\, (d-3)^2}  +
   \left({2\, A_1 \over (d-3)\, (1-y)} \right)^{2 \over d-3}\, d\Omega^2_{d-2}
   \] \nonumber
   \eea
  which is one step away from the standard form (\ref{Schw}) through
the use of the transformation rules (\ref{x-to-r}).

Let us study the various regions in the coordinate $x$ using
(\ref{x-to-r}). $x=0$ is a singularity and it separates the $x$ axis
into two regions: positive and negative $x$. For small $x$ on either
side of $x=0$ ($A_1\, |x| \ll 1$) $r^{d-3}=[(d-3)|x|]^{-1}$ and we
get flat space (Minkowski). As $x \to -\infty$ $g_{tt} \to 0$ and we
reach a horizon, and thus negative $x$ represents the outer region
of the black hole. More precisely, for $x \ll -1$, $x \propto
\log(r-r_0)$ and as such is asymptotically proportional to the
standard ``tortoise'' coordinate $r^*$ (defined by $dr^*=dr/f(r)$).
As $x \to +\infty$ $g_{tt} \to \infty$ (with no intermediate
horizon) and thus positive $x$ describes a negative mass black hole.
Finally, the inside of the black hole is not seen in these
coordinates, since it requires a changes of signature for both
$dx^2,\, dt^2$. If one were to perform this change, the potential
term in (\ref{action}) would receive a minus sign, and that would
change us from case (b) to case (c) in figure \ref{exp-pot-figure},
changing accordingly all $\sinh$'s into $\cosh$'s in the formulae
(\ref{x-metric},\ref{4d}).

\presub {\bf Generalization and the charged black hole}.
The derivation above generalizes to any spherical black hole or
black brane, the essential feature being for the metric to be of
co-homogeneity 1. Let us demonstrate it for the case of the charged
black hole (Reissner-Nordstr\"{o}m).

We need to supplement the $A,B,\,C$ fields of the ansatz
(\ref{general-ansatz}) with a vector potential $N_\mu$,
\footnote{The letters $A,\phi$, which are a more standard notation,
were already used.} and since we are interested in static,
electrically charged black holes only $N_0$ is non-vanishing and so
we may denote it simply by $N$. The action is taken to be the
standard Einstein-Maxwell action\footnote{Up to an overall minus
sign, as in the \Schw case, due to the Euclidean nature of the
essential coordinate $r$.} \be
 S = -\frac{1}{16\, \pi\, G} \int dV \( R - \frac{1}{4}\, F_{\mu\nu}\,
 F^{\mu\nu} \) \ee
where $F_{\mu\nu}=\del_\mu N_\nu - \del_\nu N_\mu$. In the static
spherically symmetric case the action becomes \bea
 S_p &=& {\Omega_{d-2} \over 16 \pi G}\, \int dt\,  \int dx\,
 e^{A+B+(d-2)C} \cdot \non
 &\cdot& \[-e^{-2\,B} [2(d-2)\,A'\, C'+(d-2)(d-3)\,C'^2]
  -(d-2)(d-3)\, e^{-2\, C} - \half\, e^{-2\, A-2\, B}\, N'^2\]  \non
 \label{gauged-unfixed-action2}\eea

We note that $N$ is an ignorable field. Therefore we continue to fix
the gauge with (\ref{gauge-choice}) such that the kinetic term for
$A,C$ will be canonical. The gauge-fixed action\footnote{After
pulling out some constants as in the \Schw case.} is \be
 S = \int dx \( \hA^2 - G^2 -e^{2\, G} - \half e^{-2 \hA}\, N'^2
 \) \ee
where we define a shifted $A$ by \be
 \hA:= A- \half \log{\frac{d-3}{d-2}} \ee
 and the constraint is \be
 0=-\left. {\del S_p \over \del B} \right|_{B=A+(d-2)C} =
  \hA^2 + G^2 -e^{2\, G} - \half e^{-2 \hA}\, N'^2 ~. \label{constr-RN} \ee

Since $N$ is an ignorable field, it is convenient at this point to
transform into its conjugate momentum \be
 Q = -e^{-2 \hA}\, N' ~. \label{N-EOM} \ee
$Q$ is conserved, and this can be considered a first integral of
$N$'s equation of motion. A partial Legendre transform of $S$ with
respect to $N$ \footnote{Such a function is called a ``Routhian''
 \cite{LL-mechanics-s41}.} yields after a multiplication by an overall minus sign \be
 R = \hA^2 - G^2 -e^{2\, G} + \half e^{+2 \hA}\, Q^2 \label{Routhian}
 \ee
This action (\ref{Routhian}) decouples with respect to $\hA,G$,
except for the constraint (\ref{constr-RN}). The latter now reads
\be
 E_A = E_G \label{constr-RN2} \ee
 where we naturally define the $A,G$ energies to be \bea
 E_A &:=& \hA'^2 - \half e^{2 \hA}\, Q^2 \non
 E_G &:=& G'^2 - e^{2\, G} \eea
Here we shall solve the equations of motion explicitly for the case
of positive energies \be
 E_A = E_G =A_1^{~2} \ee
 which will be seen to imply that the black hole is within
the extremality limit.

The solution to the equations of motion is given by \bea
 e^{-\hA} &:=& \frac{Q}{\sqrt{2}\, A_1}\, \Big| \sinh \( A_1(x-x_A) \) \Big|
 \non
e^{-G} &:=& \frac{1}{A_1}\, \Big| \sinh (A_1 x) \Big|
\label{RN-soln} \eea
 where $x_A$ is an arbitrary integration constant, and the analogous
constant $x_G$ was set to zero without loss of generality due to
invariance under $x$ translations. $N$ can be integrated from
(\ref{N-EOM})
 \be
 N = \frac{2\, A_1}{Q}\, \coth \( A_1(x-x_A) \) +N_0 ~.\ee

The standard Reissner-Nordstr\"{o}m solution
\cite{Reissner,Nordstrom,Tangherlini} is given by \bea
 ds^2 &=& -f\, dt^2 + f^{-1}\, dr^2 + r^2\, d\Omega^2_{d-2} \non
 f &=&1-{2 m \over r^{d-3}} + {q^2 \over r^{2(d-3)}} \non
 N &=& c\, {q \over r^{d-3}} +N_0  \label{RN} \eea
 where $m,q$ are related to the mass and charge through some
constants and in the current conventions $c=\sqrt{2(d-2)/(d-3)}$.

The relation between $x$ and the standard $r$ coordinate is given by
\bea
 \frac{1}{r^{d-3}} &=& e^{A-\hG}= \frac{\sqrt{(d-3)^3}}{\sqrt{d-2}}\,
 e^{\hA-G} = \frac{\sqrt{2(d-3)^3}}{\sqrt{(d-2)}}\,
 \frac{1}{Q}\,  \left| \frac{\sinh (A_1
 x)}{\sinh \( A_1(x-x_A)\)} \right| = \non
 &=& \frac{\sqrt{2(d-3)^3}}{\sqrt{d-2}}\, \frac{|\sinh (A_1\, x_A)|}{Q}\,
  \Big| \coth\( A_1(x-x_A)\) + \coth( A_1\, x_A) \Big| \eea
Let us map the various ranges of $x$ and $r$: assuming without loss
of generality that $x_A \ge 0$ we have that while $-\infty \le x \le
0$ $r$ is in the range $r_+ \le r \le +\infty$, while $0 \le x \le
x_A$ $r$ ranges over $+\infty \ge r \ge 0$ (a negative mass naked
singularity), and while $x_A \le x \le +\infty$ $r$ is in the range
$0 \le r \le r_-$.

The relations between the 3 parameters $Q,A_1,x_A$ (\ref{RN-soln})
and the 2 standard parameters $q,m$ of (\ref{RN}) can be read by
comparing the form of $g_{tt}$ \be
 -g_{tt}=e^{2 A}= \(\frac{A_1}{(d-3)\, \tQ\, \sinh{(A_1\, x_A)}}\)^2\,
  \[ 1-2\, \cosh{(A_1\, x_A)} \frac{\tQ}{r^{d-3}} +
 \(\frac{\tQ}{r^{d-3}}\)^2 \] \ee
 where $\tQ=\sqrt{(d-2)/(2(d-3)^3)}\, Q$. In the standard form \be
1=\sqrt{-g}_{tt}|_{r \to \infty}=A_1/\( (d-3) \tQ \sinh{(A_1\,
x_A)}\) \ee and then by comparison \bea
 q &=&\tQ  \non
 m &=& \cosh{(A_1\, x_A)}\, \tQ ~.\eea
 In addition, we can use the third parameter to change
the gravitational potential at infinity $g_{tt}|_{r \to \infty}$.

\presub {\bf Open question}. It would be interesting to see if
similar action techniques can be of use in deriving black hole
metrics with a higher dimension of co-homogeneity such as the
co-homogeneity 2 rotating Kerr black hole.

\presub {\bf Note added}. In response to several messages which I
received I would like to clarify a few points and relate to previous
work.

The original motivation was to find an optimal analytical gauge to
calculate the negative mode of the \Schw black hole \cite{nGPY}, and
the current derivation appeared as a spin-off.

This derivation rests on the following two ideas: an action
approach for a ``maximally general ansatz'' and a specific
action-motivated gauge-fixing. This gauge fixing produces the
current form of the metric (\ref{x-metric}). None of these ideas
is revolutionary: the action approach is not foreign to GR ever
since the days of Hilbert \cite{Hilbert} and Weyl
\cite{Weyl},\footnote{See \cite{DeserFranklin} for a recent action
treatment of Schwarzschild and references therein.} if not as
common as in other branches of physics; and while the gauge fix is
natural from the action perspective it is not impossible that
another gauge will prove ``equally natural''. Still, I found the
combination appealing and since I had not seen such a derivation
before, I decided to advertise it on the archives, and perhaps
motivate further research along these lines.

I was informed that the current form of the metric (\ref{x-metric})
did appear already in essence in \cite{Cavaglia}. While the methods
of derivation share similarities being both action-based, they are
still quite different.
Our second form of the metric (\ref{y-metric}) is closely related to
\cite{HeadWise}, though the methods differ. Altogether however,
judging by the correspondence which I received so far, it could very
well be that the current derivation did not appear yet.

\vspace{0.5cm} \noindent {\bf Acknowledgements}

This research is supported in part by The Israel Science Foundation
grant no 607/05 and by the Binational Science Foundation
BSF-2004117.

\end{document}